\documentclass[lettersize,journal]{IEEEtran}
\usepackage{amsmath,amsfonts}
\usepackage{algorithmic}
\usepackage{algorithm}
\usepackage{array}
\usepackage[caption=false,font=normalsize,labelfont=sf,textfont=sf]{subfig}
\usepackage{textcomp}
\usepackage{stfloats}
\usepackage{url}
\usepackage{verbatim}
\usepackage{graphicx}
\usepackage{cite}
\usepackage{url}
\usepackage{multirow}
\usepackage{xcolor}
\usepackage{svg}
\usepackage{threeparttable}

\usepackage{xspace}

\begin{document}
\title{Revolutionizing Wireless Communications with Space Data Centers: Applications and Open Challenges}

\author{Minghao~Sun,
        Zehui~Chen,
        Jinbo~Hou,
        Kezhi~Wang,~\IEEEmembership{Senior Member, IEEE,}
        Xiaoli~Chu,~\IEEEmembership{Senior Member, IEEE,}
        }


\maketitle

\begin{abstract}
Space data centers (SDCs) are emerging as a promising orbital computing infrastructure for the future AI industry. Unlike conventional satellites that mainly serve as relay nodes or lightweight onboard processors, SDCs integrate communication, computing, storage, and control capabilities in orbit, enabling persistent service support for data-intensive and intelligence-driven space applications. 
In this article, we investigate how SDCs may transform space communication paradigms from connectivity-oriented data transmission toward task-oriented and service-centric information exchange.
We first present a hierarchical SDC network architecture consisting of access, relay, computing, and control layers, and outline possible deployment strategies. 
We then explore representative future application scenarios enabled by SDCs, highlighting their communication characteristics and associated research challenges. Simulation results further demonstrate the effectiveness of SDCs in reducing control-layer latency in hierarchical space networks.
Finally, we identify key research directions toward the practical deployment of SDCs.
\end{abstract} 


\section{Introduction}

\IEEEPARstart{I}{n} recent years, the rapid advancement of artificial intelligence has significantly increased the computing demand on terrestrial data centers \cite{AI_SDC}. The operating power of modern hyperscale data centers has approached 100 megawatt (MW) and is expected to scale further toward gigawatt (GW)-class infrastructure, placing increasing pressure on regional power grids, site availability, and cooling systems \cite{Starcloud_wp}.
In contrast, the space environment offers a more favourable resource foundation for large-scale computing platforms \cite{SDC_survey}. 
StarCloud has proposed an orbital data center powered by ultra-large solar arrays spanning more than four kilometers, targeting multi-gigawatt-level power generation. The company has also demonstrated early feasibility through the launch of a prototype computing platform equipped with H100 GPUs \cite{Starcloud1_H100}.
Against this background, the Space Data Center (SDC) is evolving from a conceptual vision into an active engineering endeavour. At the same time, space systems are also undergoing rapid transformation. 
Emerging applications, including mega-scale remote sensing constellations, global IoT access, in-orbit artificial intelligence, digital twins, and ground-space collaborative computing, are continuously increasing both the data generation rate and task complexity within space networks. 
Meanwhile, the growing demand for in-orbit manufacturing, maintenance, and servicing is expected to generate massive volumes of operational data and new requirements for intelligent processing \cite{orbit_eng}.

However, most existing satellite communication systems still treat satellites primarily as relay nodes or lightweight onboard processing platforms, making them inadequate for supporting the sustained computing, storage, and service coordination required by future workloads \cite{SDC_com}.
Accordingly, the significance of SDCs extends beyond alleviating pressure on terrestrial infrastructures or supporting future orbital services, they also have the potential to enable a new class of communication scenarios. Future SDC communication is therefore not simply a matter of transporting more data to orbit, but of enabling task-oriented and service-centric information exchange across ground-space networks.
This challenge is particularly important because space artificial intelligence training and model updating are inherently communication-intensive, while orbital networks still remain substantially weaker than terrestrial data center networks in terms of bandwidth, connection continuity, and latency stability. 
For this reason, understanding the future communication scenarios, architectural design, and key bottlenecks of SDCs has become central to translating the concept into practical deployment.

This article provides an architectural and application-oriented perspective on SDCs as an emerging communication and computing paradigm for future space networks. 
We first examine how SDCs may reshape future space communication scenarios, present a hierarchical SDC network architecture that comprises access, relay, computing, and control layers, and outline possible deployment strategies. 
The main contribution of this article is the proposed SDC network architecture and the identification of emerging application scenarios enabled by SDCs. For each scenario, we analyze its key characteristics and discuss the associated research problems.

\begin{figure*}[t]
    \centering
    \includegraphics[width=1.8\columnwidth]{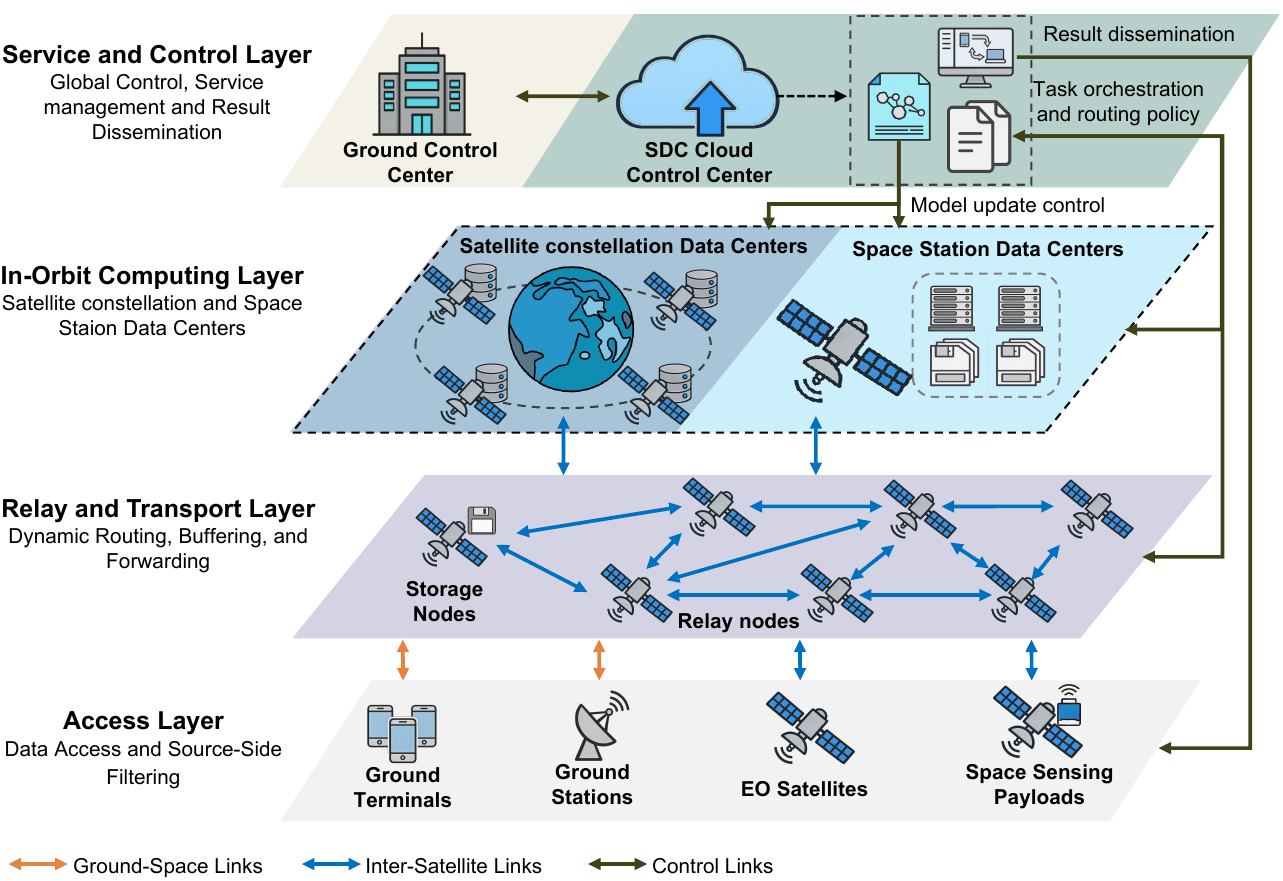}
    \caption{Proposed hierarchical SDC network architecture}
    \label{framework}
\end{figure*}

\section{SDC Network Architecture}

In this section, we first explain why SDCs are becoming a systemic requirement for future space communication systems, and then present a hierarchical SDC framework that integrates communication, computing, storage, and control. We further discuss the potential deployment of SDCs from the perspectives of energy supply, thermal management, and orbital configuration.

\subsection{Toward a New SDC-Centered Communication Paradigm}

With the continuous evolution of space information infrastructures, the communication requirements of SDCs are becoming increasingly diverse, which is driven by fundamental changes in the communication entities, task patterns, and service logic of space networks.

The scale of data generated within space networks is increasing rapidly. Future mega-scale Earth observation (EO) constellations, continuous video monitoring payloads, distributed sensor networks, and globally connected space IoT systems will continuously generate massive volumes of heterogeneous data in orbit \cite{link_data}. 
If the traditional “data acquisition–raw downlink–centralized ground processing” paradigm persists, ground-space links will become a primary bottleneck, due to limited capacity, intermittent visibility windows, and insufficient connection continuity to support the growing demand for data return. 
In this context, simply increasing transmission capability is not enough to fundamentally solve the problem. 
Instead, in-orbit hub nodes are required to support data caching, pre-processing, filtering, fusion, and task scheduling, thereby reducing redundant data transmission and improving the efficiency of useful information delivery.
Meanwhile, advances in in-orbit artificial intelligence, digital twins, cooperative sensing, and autonomous network control are enabling future space networks to support increasingly sophisticated workloads, such as pretrained model deployment, model fine-tuning and updating, target event filtering, distributed state synchronization, and cross-node collaborative decision-making \cite{SDC_AI_train,digital_twin}. 
However, if the space network lacks core nodes with task processing and coordinated scheduling capabilities, many of these advanced services and decisions would still need to be offloaded to ground infrastructures for execution, resulting in longer transmission paths, higher resource overhead, and lower system responsiveness. Conventional architectures that treat satellites as relay nodes or lightweight processing platforms are no longer adequate to support such persistent, complex, and task-oriented communication demands, which are tightly coupled with computing, storage, and control. In contrast, SDCs provide a continuously available in-orbit service entity with relatively strong computing and storage capabilities and the ability to support complex service logic, thereby driving space networks from mere connectivity infrastructure toward a multifunctional integrated service platform.

Overall, SDCs introduce a new organizing paradigm for future space communication systems. 
By aggregating distributed data flows in space, accommodating complex task flows, supporting continuous exchange of model and state flows, and enabling coordinated information processing across constellations, layers, and regions, SDCs provide a foundational platform for large-scale space network intelligence and service orchestration. 
This implies that future SDC research should extend beyond traditional considerations such as energy supply, computing capability, and thermal management, and pay greater attention to how SDCs reshape information organization, resource scheduling, and service delivery mechanisms in space communication systems.

\subsection{A Hierarchical SDC Framework} 

As shown in Fig. \ref{framework}, we propose a multi-layer orbital computing architecture comprising the access layer, relay and transport layer, in-orbit computing layer, and service and control layer, which together enable the integrated operation of communication, computing, storage, and control functions.

The access layer consists of ground terminals, ground stations, EO satellites, and various sensing payloads and is responsible for data acquisition, source-level filtering, and preliminary task classification.
Above the access layer, the relay and transport layer serves as the dynamic communication backbone of the SDC system. It consists of relay satellites, inter-satellite links, dynamic routing modules, and storage nodes equipped with buffer-assisted forwarding mechanisms. 
Since data sources cannot always maintain direct connectivity with SDC nodes, this layer forwards task data through multi-hop satellite paths while adaptively adjusting routing decisions according to link availability, satellite mobility, traffic load, and service priority. 
Inter-satellite links provide high-capacity transmission among satellites, while buffer-assisted forwarding at storage nodes helps mitigate the impact of intermittent connectivity and temporary link interruptions. 
To support the high-capacity and low-latency inter-satellite communications, this layer can employ laser communication technologies, with the capacity of inter-satellite optical links already reaching 400 Gbps \cite{laser}. As a result, the relay and transport layer acts as a critical bridge between distributed access nodes and centralized or distributed orbital computing resources, ensuring efficient and reliable delivery of data, control messages, and computation results across time-varying space networks.

The in-orbit computing layer forms the core of the SDC system. As shown in Fig.\ref{framework}, this layer may adopt two representative architectural forms: satellite-constellation data centers and space-station data centers. 
In the satellite-constellation architecture, multiple computing satellites are interconnected through inter-satellite links to form a distributed orbital computing platform. This architecture improves coverage, scalability, and service continuity, and allows computing resources to be expanded by adding new nodes. In the space-station data center architecture, high-performance servers can be deployed in a relatively centralized platform, which may simplify maintenance, resource management, and large-scale computing integration.  
When required, this layer can further evolve into a coordinated multi-SDC system, where computing tasks are partitioned, migrated, or jointly executed among several orbital data center nodes.
In addition to primary SDC nodes equipped with substantial computing, storage, and communication capabilities, auxiliary computing and storage satellites can be deployed to support collaborative processing, caching, backup, and load balancing.

Finally, the service and control layer serves as the interface between the orbital computing infrastructure and ground-based management and cloud platforms. It provides global control, service management, task orchestration, model update management, and result dissemination. Through the ground control center and SDC cloud control center, this layer monitors the status of satellites, computing nodes, storage resources, and communication links, and generates scheduling decisions for task allocation, routing, resource provisioning, and service recovery. It also supports AI model lifecycle management, including model uploading, version synchronization, retraining coordination, and update validation. 
Once in-orbit computation is completed, the service and control layer coordinates the dissemination of results to ground users, cloud platforms, or other satellite nodes.

Under this framework, the distributed architecture formed by core SDC nodes, auxiliary computing nodes, and storage nodes enhances both system resilience and scalability. Non-real-time tasks can be offloaded to auxiliary computing nodes, while storage nodes support data caching, intermediate state preservation, and model hosting, thereby reducing the processing burden on central nodes. Meanwhile, dynamic interconnections among satellites improve system evolvability, enabling incremental integration of new nodes and seamless replacement of aging platforms without interrupting ongoing services.
However, this modular and scalable architecture also introduces additional overhead in data movement, node coordination, and service orchestration. In highly dynamic space communication environments, it further brings new challenges in cross-layer coordination and system design.
\begin{figure}[t]
    \centering
    \includegraphics[width=1\columnwidth]{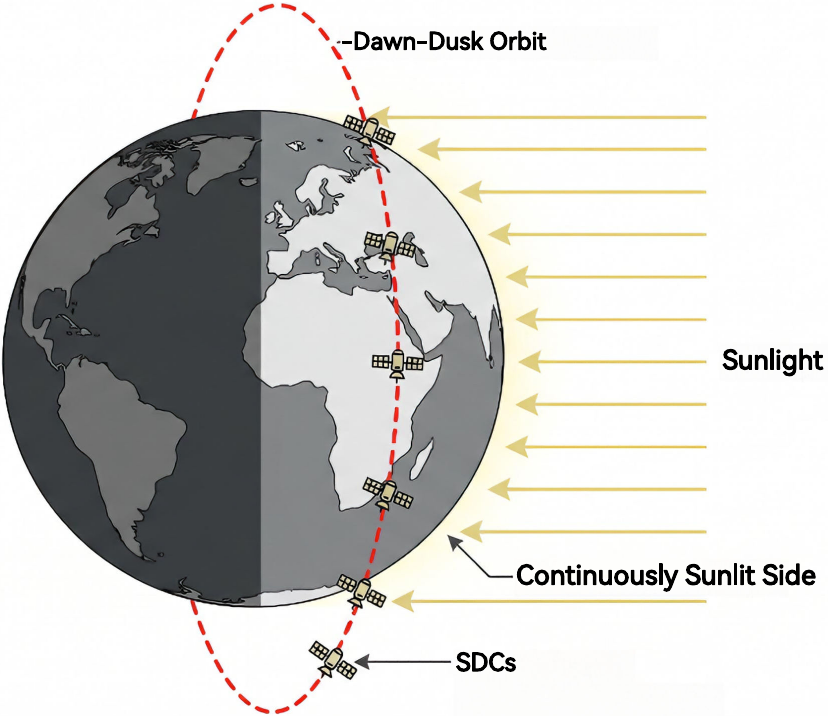}
    \caption{Dawn-dusk sun-synchronous orbits}
    \label{orbit}
\end{figure}

\subsection{Engineering Considerations and Orbital Deployment}

At present, the high energy consumption of data centers has become one of the major bottlenecks limiting their large-scale expansion. 
According to \cite{energy_report}, approximately 58\% of energy consumption in terrestrial data centers is attributed to cooling. 
In contrast, the space environment lacks convective heat transfer, meaning that waste heat generated by SDCs cannot be removed through air cooling and must instead be dissipated via large-area radiators through infrared radiation to deep space. Nevertheless, with appropriate thermal design and the benefit of the low background temperature in orbit, radiative cooling can significantly reduce the energy burden associated with conventional active cooling systems.
Meanwhile, given that SDCs typically require large-scale solar arrays to ensure a sustained power supply, the radiator surfaces can, subject to thermal-control, attitude, and structural constraints, be integrally designed with the anti-sun side of solar panels or other shadowed structures, thereby improving the utilization efficiency of platform surface area and reducing the need for additional deployable thermal structures \cite{SDC_mag}.
To further improve power supply stability, SDCs may be deployed in dawn-dusk sun-synchronous orbits, as depicted in Fig.\ref{orbit}. Such orbits allow the orbital plane to remain close to the day-night terminator for extended periods, thereby maximizing solar exposure, reducing eclipse duration, and improving the continuity of solar energy harvesting \cite{google_orbit}. 
Meanwhile, satellites can maintain tight formations with relative distances ranging from hundreds of meters to several kilometers through precise formation control, thereby reducing signal attenuation caused by long communication distances and improving inter-satellite communication efficiency.

\section{Emerging Communication Scenarios for SDC-Enabled Applications}

\begin{figure*}[t]
    \centering
    \includegraphics[width=2\columnwidth]{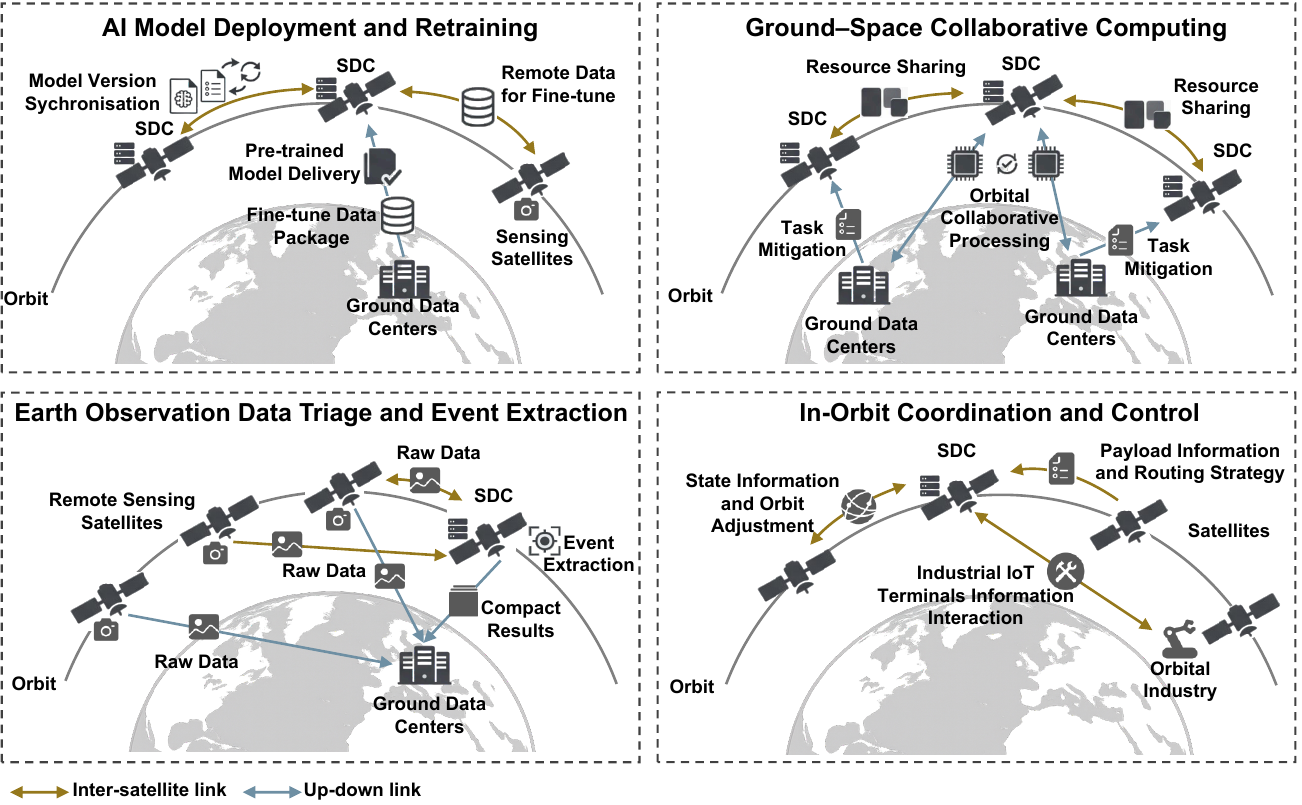}
    \caption{Four representative communication scenarios enabled by SDCs}
    \label{application}
\end{figure*}

In this section, we examine representative communication scenarios enabled by SDCs, as depicted in Fig.\ref{application}. We further analyze the communication characteristics and key requirements of each scenario, and validate the latency advantage of an SDC-centered control architecture through a multi-constellation simulation.

\subsection{Representative Future Communication Scenarios Enabled by SDCs}

\textbf{AI Model Deployment and Retraining:}
AI model training and retraining represent one of the core functions of SDCs. However, SDCs are generally not well-suited for training entirely new large-scale models from scratch, especially when the training data are mainly stored on the ground. 
The main bottleneck lies not only in the limited onboard computing capability, but also in the substantial communication cost of transferring massive datasets from terrestrial infrastructures to orbit. Uploading full raw training datasets would consume scarce ground-space link resources, introduce long transmission delays, and incur high energy and scheduling overhead.
Therefore, compared with full-scale model pre-training, AI-related workloads supported by SDCs can be more reasonably divided into two categories. The first is in-orbit deployment of pretrained models uploaded from the ground, where the main communication task is to reliably deliver large model files, configuration parameters, and execution environments to the SDC. 
For example, in November 2025, Zhijiang Laboratory in China uploaded the Tongyi Qianwen (Qwen3) large language model to satellites launched under the Three-Body Computing Constellation program.
The second category involves the retraining, incremental updating, or fine-tuning of models already deployed in orbit. In this case, communication mainly involves the transmission of selected samples, model updates, gradients, adapters, metadata, and validation feedback among ground nodes, relay satellites, and SDC platforms.
These tasks are usually latency-tolerant. Their communication design is therefore not driven by ultra-low latency requirements, but by the need to ensure transmission accuracy, model integrity, update consistency, privacy preservation, security protection, and reliable task completion under intermittent connectivity. 

For large AI model deployment, even minor corruption or loss of parameter data may result in deployment failure or degraded inference performance.
Therefore, accurate model injection is a fundamental requirement.
Since ground-space links may be available only during limited visibility windows, model files may need to be segmented, compressed, encoded, and transmitted across multiple contact periods. Model updates and training-related data should be scheduled according to link capacity and task priority, while packet-loss recovery, model-integrity verification, and execution-environment validation mechanisms are required.
For model retraining and fine-tuning, inconsistent updates across orbital nodes may lead to model-version divergence, unstable service behavior, or unreliable decision outputs. Effective synchronization, version control, and rollback strategies are therefore essential. 
In addition, training data, model parameters, gradients, and intermediate representations may contain sensitive information, making them vulnerable to eavesdropping, tampering, and adversarial attacks.
Hence, secure and privacy-preserving transmission must be maintained throughout the model lifecycle.

\textbf{Ground–Space Collaborative Computing:}
When ground data centers operate under heavy load or face constraints in power supply and thermal management, selected delay-tolerant yet computation-intensive workloads can be offloaded to SDCs to relieve pressure on ground-based computing and energy resources. 
This service model is particularly suitable for tasks that require substantial computing capability but do not demand real-time or interactive responses.
In addition, because SDCs can connect with multiple geographically distributed ground data centers through satellite networks, they may serve as orbital coordination hubs for cross-regional task migration, resource sharing, and collaborative processing. This capability is especially valuable when terrestrial backbone networks are congested, damaged, or unevenly provisioned, thereby enhancing the overall resilience and coordination capability of multi-data-center systems.

In this scenario, a prominent communication characteristic is that task uploading, intermediate data transfer, model synchronization, and result downloading may generate short-term bursty traffic, especially when multiple ground data centers migrate workloads to orbital platforms at the same time. If not properly managed, such traffic may cause congestion not only on ground-space access links, but also along inter-satellite relay paths and at SDC ingress buffers.
Therefore, SDCs must coordinate link resources, onboard storage, buffering capacity, and scheduling policies across the satellite network. Burst-aware transmission and buffering mechanisms are required to handle sudden workload migration, model synchronization, and result-return traffic, thereby improving the continuity, stability, and efficiency of large-scale data transfer under dynamic network topologies.
Meanwhile, efficient task allocation and cross-regional resource management mechanisms are needed to jointly optimize communication and computing resources. Such mechanisms should determine which workloads are best processed on the ground, which are suitable for offloading to SDCs, and when migration and execution should be performed, particularly when SDCs simultaneously serve multiple ground data centers with different service priorities and network conditions. 
For example, workloads with small input sizes but high computational complexity may be well suited for SDC processing, whereas data-intensive workloads requiring frequent interaction with ground-based databases are generally more efficiently executed on terrestrial infrastructures.

\textbf{Earth Observation Data Triage and Event Extraction:}
Conventional onboard remote sensing has long been constrained by the limited computing, storage, and energy resources of individual satellites. In traditional satellite systems, most remote sensing payloads mainly perform data acquisition and basic preprocessing, while computation-intensive operations, such as high-resolution image interpretation, target detection, change detection, disaster assessment, and multi-source data fusion, are typically executed only after the raw data have been downlinked to terrestrial data centers \cite{remote}. 
As the spatial, spectral, and temporal resolutions of remote sensing systems continue to improve, this processing paradigm is becoming increasingly inefficient. 
Meanwhile, large-scale Earth observation constellations continuously generate massive volumes of imagery and video streams, yet ground station access remains intermittent and geographically limited. 
As a result, data may accumulate onboard during non-visible periods, creating substantial storage pressure and leading to bursty downlink traffic once connectivity with ground infrastructure is restored.
SDCs can effectively alleviate these bottlenecks by providing substantially greater in-orbit computing and storage resources. 
In an SDC-enabled remote sensing architecture, selected sensing data can be offloaded from observation satellites to orbital computing or storage nodes through inter-satellite links. The SDC can then perform advanced onboard processing, including image enhancement, feature extraction, target recognition, anomaly detection, data compression, and task-oriented information filtering. This architecture makes it feasible to deploy more advanced machine learning models with higher computational demand in orbit, particularly for applications requiring timely situational awareness, such as wildfire monitoring, flood assessment, maritime surveillance, emergency response, and military- or security-related observation. By extracting task-relevant information before downlink, SDCs can reduce the volume of raw data transmitted to the ground while improving the responsiveness of remote sensing services.

However, this does not imply that all remote sensing data should be fully processed in orbit before being transmitted to the ground. Since the computing, storage, inter-satellite link, and downlink resources of SDCs are typically shared across multiple tasks and satellites, and may even be jointly used by different organizations or agencies, these resources must be managed in an integrated manner. If all raw images and video streams are forwarded to the SDC without selection, congestion may first emerge within the inter-satellite network. Likewise, if all processing results and intermediate data are transmitted to the ground without priority-aware scheduling, the downlink bottleneck is merely shifted rather than resolved. Therefore, SDC-enabled remote sensing requires a more selective, task-aware, and resource-coordinated communication and computing framework. Such a framework should jointly consider data volume, task urgency, model complexity, and link availability to determine which data should be processed in orbit, which should be compressed or filtered at the feature level, which should be temporarily buffered, and which should be directly downlinked to ground stations. For example, time-sensitive monitoring tasks may require rapid feature extraction and result dissemination, whereas scientific observation data may tolerate longer delays but demand high-fidelity preservation. Furthermore, when an orbital data center serves satellites belonging to multiple agencies or commercial operators, additional issues related to fairness, isolation, service-level assurance, and security in multi-tenant resource management must also be addressed.

\textbf{In-Orbit Coordination and Control for Large-Scale Space Systems:}
SDCs can provide low-latency services for large-scale satellite systems in coordinated orbital networks. As satellite constellations expand from tens or hundreds of spacecraft into larger multi-layer systems, the complexity of constellation management increases significantly. Through inter-satellite links, an SDC can rapidly aggregate operational information from multiple satellites and leverage its relatively strong onboard computing capability to execute advanced network coordination and optimization algorithms, thereby serving as a central coordination hub for constellation-level decision-making in multi-satellite networks. 
With the support of SDCs, the system can not only dynamically allocate inter-satellite link resources, adjust data forwarding paths, coordinate observation tasks among Earth observation satellites, and optimize the assignment of computing tasks across orbital nodes, but also support multiple satellites in jointly performing wide-area sensing, distributed beamforming, and cooperative data collection, thereby enabling global scheduling, resource orchestration, and mission-level decision making. 

Compared with the traditional “ground download–process–return” loop, this in-orbit control mode can significantly reduce decision latency, improve system-level responsiveness, provide near-real-time coordination support, and relieve the computing burden on resource-constrained satellites.
This capability is also highly relevant to emerging space industrial scenarios, including in-orbit servicing, debris removal, in-orbit assembly, and space manufacturing \cite{NASA_2026}. 
Such applications typically involve the coordinated operation of multiple robotic spacecraft, servicing vehicles, inspection satellites, storage modules, and manufacturing platforms within a shared orbital environment. Their operation depends heavily on accurate state awareness, continuous task coordination, reliable control signaling, and rapid response to unexpected events. In this context, SDCs can further serve as data and control centers for future space industrial infrastructure, supporting large-scale autonomous robotic systems in tasks such as component replacement, equipment inspection, and on-orbit maintenance.

In future scenarios, SDCs are expected to evolve into integrated intelligent coordination centers spanning communication networking, distributed computing, autonomous control, and space system security, which will introduce new communication and control challenges. 
Highly coordinated tasks require low-latency inter-satellite connectivity. Unlike conventional payload data, control information is typically small in volume but highly sensitive to latency, packet loss, and out-of-order delivery. 
Therefore, routing, link selection, and resource allocation mechanisms must ensure that control messages are prioritized and delivered reliably within strict timing constraints under time-varying orbital topologies, while avoiding excessive degradation of task-data throughput.
Moreover, the SDC control layer continuously exchanges large volumes of control information related to satellite network operation, space IoT coordination, and system state management.
In such a context, Faults, cyberattacks, command delays, or corrupted state information may trigger system-wide impacts. 
As a result, control-layer communications in SDCs must provide high reliability as well as strong resistance against interference and eavesdropping. 
Furthermore, when multiple operators or agencies share the same space-ground communication and coordination infrastructure, issues such as access control, trust management, task isolation, and differentiated security assurance in multi-tenant or multi-agency satellite systems become equally critical.

\subsection{Validation for In-Orbit Control Center}
\begin{figure}[t]
    \centering
    \includegraphics[width=1\columnwidth]{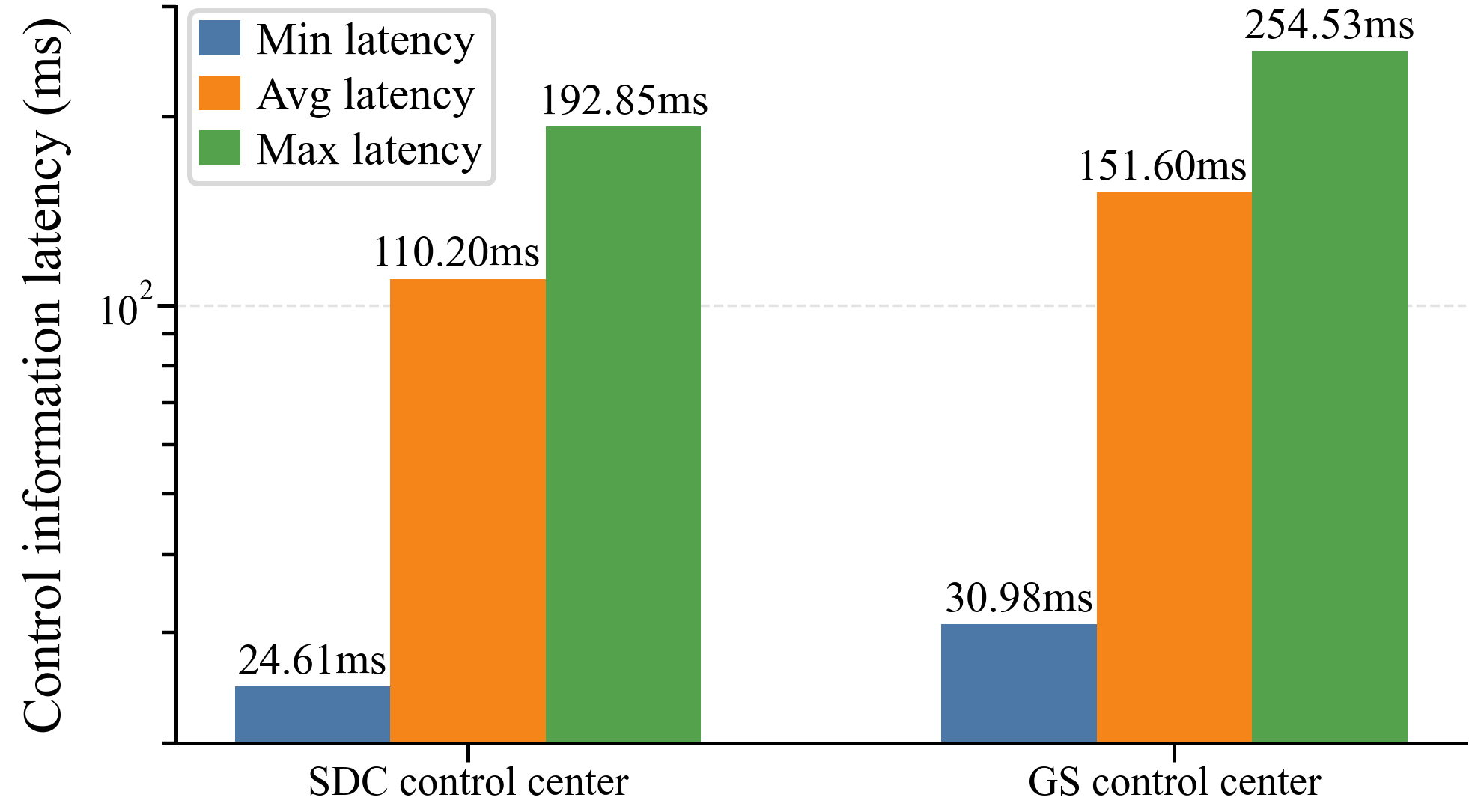}
    \caption{Control information latency between SDC-centered and GS-centered strategies}
    \label{result}
\end{figure}

In Fig.\ref{result}, we simulate the control-layer information latency in a multi-constellation LEO satellite network at an altitude of 550 km. 
In the simulated network, each constellation consists of eight satellites, with four constellations considered in total. Each satellite periodically generates operational state information, including position, velocity, computing load, and other status indicators, with a total data size of 10 kB.
The state information is transmitted through multi-hop paths to the control center, where it is processed and subsequently returned to the satellites. Two control architectures are compared: one using the SDC as the in-orbit control center, and the other using a ground station as the control center. The access delay for each multi-hop transmission is set to 2 ms, and the processing delay at the control center is set to 20 ms.
The results show that the SDC-centered space control architecture achieves lower control-layer latency, demonstrating improved timeliness for low-latency coordination in hierarchical satellite networks.

\section{Future Research Directions}

Future research on SDCs should jointly address high-capacity communications, cross-layer resource orchestration, and secure service delivery, thereby driving SDCs from a promising in-orbit computing concept toward an integrated intelligent center for computing and control coordination in future space networks. From this perspective, this section discusses the open challenges introduced by SDCs through the lens of three core system bottlenecks.

First, communication capacity remains the fundamental bottleneck in representative scenarios such as model deployment and retraining, ground–space collaborative computing, EO data triage and event extraction, and large-scale in-orbit coordination and control.
A substantial gap still exists between the capabilities of current ground–space and inter-satellite links and the requirements of data-center-scale services. 
Addressing this gap requires not only increasing link capacity, but also improving effective information delivery and reducing communication overhead, for example, through spatial multiplexing and semantic communications.

Second, because different application scenarios impose different scheduling requirements, future work should develop software-defined scheduling frameworks with cross-layer coordination capability. 
Such frameworks should jointly optimize communication resources, elastic buffering, task prioritisation, state synchronization, and service continuity, thereby enabling efficient operation under asynchronous and dynamic orbital environments.

Finally, as SDCs evolve into key nodes in future space control networks, security will shift from a secondary concern to a primary design objective. 
In particular, further research is needed on protecting sensitive control information, ensuring multi-tenant data privacy, enforcing access control, enabling task isolation, and strengthening trust management across different stakeholders.

\section{Conclusion}

In conclusion, this article has presented SDCs as an emerging orbital infrastructure that can shift future space networks from simple data forwarding to task-oriented and service-centric communication. 
To support this vision, a hierarchical SDC  network architecture has been proposed. 
Based on this framework, four representative scenarios have been discussed, demonstrating that SDCs can support data processing, model management, workload migration, event-aware sensing, and low-latency control closer to where space data and decisions are generated. Simulation results further demonstrate the potential latency advantage of using SDCs as in-orbit control centers. To realize this vision, future efforts should address communication capacity enhancement, cross-layer orchestration, and secure service delivery.

\section*{Acknowledgment}
This work was supported in part by the Horizon Europe Research and Innovation Program under grants 101086219 and 101086228, the UK EPSRC grants EP/X038971/1 and EP/Y028031/1, Innovate UK COMET project (No: 10099265), and Royal Society Industry Fellowship (IF$\setminus$R2$\setminus$23200104).
Xiaoli Chu is the corresponding author.

\bibliography{ref}

@article{SDC_survey,
  title={Towards Space-Based Computing Infrastructure Network: Development Trends, Network Architecture, Challenges Analysis, and Key Technologies},
  author={Kuang, Linling and Sun, Jiachen and Zhang, Jin and Cui, Huanxi and Liu, Kai},
  journal={arXiv preprint arXiv:2503.06521},
  year={2025}
}

@misc{Starcloud1_H100,
  author       = {{Starcloud}},
  title        = {{Starcloud-1: The First Satellite with an NVIDIA H100 GPU in Orbit}},
  year         = {2025},
  howpublished = {\url{https://www.starcloud.com/starcloud-1}},
  note         = {Accessed: May 19, 2026}
}

@article{link_data,
  title={Compressed robust transmission for remote sensing services in space information networks},
  author={Lu, Hancheng and Gui, Yongqiang and Jiang, Xiaoda and Wu, Feng and Chen, Chang Wen},
  journal={IEEE Wireless Communications},
  volume={26},
  number={2},
  pages={46--54},
  year={2019},
  publisher={IEEE},
  month={May}
}

@article{SDC_mag,
  title={Toward Communication-Efficient Space Data Centers: Bottlenecks, Architectures, and New Paradigms},
  author={Sun, Minghao and Chen, Zehui and Hou, Jinbo and Wang, Kezhi and Chu, Xiaoli},
  journal={arXiv preprint arXiv:2605.12681},
  year={2026}
}

@article{google_orbit,
  title={Towards a future space-based, highly scalable AI infrastructure system design},
  author={y Arcas, Blaise Ag{\"u}era and Beals, Travis and Biggs, Maria and Bloom, Jessica V and Fischbacher, Thomas and Gromov, Konstantin and K{\"o}ster, Urs and Pravahan, Rishiraj and Manyika, James},
  journal={arXiv preprint arXiv:2511.19468},
  volume={4},
  year={2025}
}

@article{orbit_eng,
  title={On-orbit servicing: Inspection repair refuel upgrade and assembly of satellites in space},
  author={Davis, Joshua P and Mayberry, John P and Penn, Jay P},
  journal={The Aerospace Corporation, report},
  volume={25},
  year={2019}
}

@article{SDC_AI_train,
  title={Advancing Earth observation: a survey on AI-powered image processing in satellites},
  author={Duggan, Aidan and Andrade, Bruno and Afli, Haithem},
  journal={European Journal of Remote Sensing},
  volume={58},
  number={1},
  pages={2567921},
  year={2025},
  month={Oct.},
  publisher={Taylor \& Francis}
}

@article{digital_twin,
  title={Twinning for Space-Air-Ground-Sea Integrated Networks: Beyond Conventional Digital Twin Towards Goal-Oriented Semantic Twin},
  author={Qiu, Yifei and Liao, Tianle and Jin, Xin and Zhang, Qinyu and Wu, Shaohua},
  journal={arXiv preprint arXiv:2512.16058},
  year={2025}
}

@misc{NASA_2026,
  author       = {{National Aeronautics and Space Administration}},
  title        = {{Rendezvous, Proximity Operations \& Docking Subsystems}},
  year         = {2026},
  howpublished = {\url{https://www.nasa.gov/reference/jsc-rendezvous-prox-ops-docking-subsystems/}},
  note         = {Accessed: 2026-05-18}
}

@article{SDC_com,
  title={Space Cloud Networks: Technologies, Architectures, and Key Enablers},
  author={Ge, Wenxiao and Fraire, Juan A and Shen, Xinxin and Zhao, Kanglian},
  journal={IEEE Communications Magazine},
  year={2026},
  publisher={IEEE},
  month={Mar.}
}

@article{remote,
  title={Link topology and multi-objective mission flow optimization for remote sensing satellites with inter-layer links and satellite-ground links},
  author={Zhong, Xiaoqing and Guo, Ningxuan and Li, Anshou and Gong, Yupeng and Wang, Yuqi and Wang, Ningyuan and Liu, Liang},
  journal={IEEE Transactions on Vehicular Technology},
  volume={73},
  number={10},
  pages={15621--15635},
  year={2024},
  month={Jun.},
  publisher={IEEE}
}

@misc{energy_report,
  title={2024 United States Data Center Energy Usage Report},
  author={Shehabi, Arman and Newkirk, Alex and Smith, Sarah J and Hubbard, Alex and Lei, Nuoa and Siddik, Md Abu Bakar and Holecek, Billie and Koomey, Jonathan and Masanet, Eric and Sartor, Dale},
  year={2024},
  month={Dec.},
  organization = {Lawrence Berkeley National Laboratory}
}

@techreport{Starcloud_wp,
  title={Why we should train AI in space},
  author={Feilden, Ezra and Oltean, Adi and Johnston, Philip},
  institution  = {StarCloud},
  type         = {White Paper},
  year={2024}
}

@article{AI_SDC,
  title={Towards a future space-based, highly scalable AI infrastructure system design},
  author={y Arcas, Blaise Ag{\"u}era and Beals, Travis and Biggs, Matthew and Bloom, Jessica V and Fischbacher, Thomas and Gromov, Konstantin and K{\"o}ster, Urs and Pravahan, Rishiraj and Manyika, James},
  journal={arXiv preprint arXiv:2511.19468},
  volume={4},
  year={2025}
}

@misc{laser,
  author       = {Andrew Jones},
  title        = {China Makes High-Speed Laser Links in Orbit},
  year         = {2025},
  month        = {may},
  day          = {12},
  howpublished = {\url{https://spectrum.ieee.org/satellite-internet-china-crosslink}},
  note         = {{IEEE Spectrum}, accessed Apr. 8, 2026}
}
\bibliographystyle{IEEEtran}  


\end{document}